\documentclass[12pt]{article}
\usepackage{amssymb,amsmath,epsfig}

\begin{document}
\title{\bf Greybody Factor for a Static Spherically Symmetric Black Hole With Non-Linear
Electrodynamics}

\author{M. Sharif \thanks{msharif.math@pu.edu.pk} and A. Raza
\thanks{aliraza008.math@gmail.com}\\
Department of Mathematics, University of the Punjab,\\
Quaid-e-Azam Campus, Lahore-54590, Pakistan.}

\date{}

\maketitle

\begin{abstract}
In this paper, we study the greybody factor for static spherically
symmetric black hole with non-linear electrodynamics. For this
purpose, we assume minimal coupling of the scalar field and find the
radial equation by using the Klein-Gordon equation. We then apply
tortoise coordinate to convert this equation into Schrodinger wave
equation which helps to find the effective potential. The behavior
of effective potential is checked for different values of the
coupling and charge parameters. We find two solutions in two
horizons named as event and cosmological horizons by using the
radial equation. We consider the intermediate regime and match these
two solutions to obtain the greybody factor and examine its behavior
graphically. It is found that the greybody factor has an inverse
relation with the coupling constant, mass, charge as well as the
radius of the black hole and has a direct relation with angular
momentum.
\end{abstract}

\section{Introduction}

Black holes (BHs) are some of the strangest and most fascinating
objects in outer space. They are extremely dense, with such strong
gravitational attraction that even light cannot escape their grasp
if it comes near enough. Some well-known BHs are developed by
Schwarzschild, Reissner-Nordstrom,  Kerr and Kerr-Newman based on
the physical parameters. Singularity is the most crucial issue in
gravitational physics as it is a point in spacetime where physical
laws break down. Bardeen \cite{1} was the pioneer in the study of
non-singular BH solutions, known as regular BHs. Kiselev \cite{2}
proposed the solutions of Schwarzschild BH surrounded by the
quintessence matter in the presence and absence of charge. Hayward
\cite{3} extended the Bardeen concept and considered the BH with a
non-singular static region of trapped surfaces.

Many BH solutions were developed for the quintessence field by using
the Kiselev algorithm. Chen and Jing \cite{4} calculated the
frequencies of the massless scalar field around static spherically
symmetric BHs. Bambi and Modesto \cite{5} applied the Newman-Janis
algorithm to Bardeen as well as Hayward BHs and obtained a family of
rotating non-singular solutions for both metrics. Xu et al \cite{6}
found a spherically symmetric BH solution by using Newman-Janis
technique and obtained a relation between perfect fluid dark matter
and the cosmological constant. They also extended this solution for
the de Sitter and anti-de Sitter spacetimes. Xu et al \cite{7}
generalized Reissner-Nordstrom BH to the Kerr-Newman-anti-de Sitter
BH and examined that there is no effect of perfect fluid dark matter
on singularity.

The laws of thermodynamics such as the law of conservation of energy
and the law of entropy (the entropy of an isolated system always
increases) also hold in BH  thermodynamics. This idea attracted
Hawking and proposed that if BH has temperature then it must emit
some radiations called Hawking radiations \cite{8}. The rate at
which it emits radiations is defined as
\begin{equation}\nonumber
\gamma(w)=\left(\frac{d^3k}{(e^{\frac{w}{t_h}} \pm 1)
8\pi^3}\right),
\end{equation}
where $t_h$, $w$ and $k$ are the Hawking temperature, frequency and
surface gravity, respectively. Here $d$ shows the change in surface
gravity from $k$ to $\delta k$. Since the emission rate has an
inverse relation to the size of BH, therefore, smaller BHs will emit
radiations faster than the larger BHs. The above expression can be
generalized for any dimension as well as massive and non-massive
particles. The positive and negative signs show boson and fermion
particles, respectively. Hawking radiations give information about
the physical features inside the BH such as charge, angular
momentum, mass, etc.

The exterior region of a BH plays the role of potential barrier for
radiations, thus the spectrum formed at the event horizon is similar
to the black-body spectrum. The potential required for Hawking
radiations to cross the curvature outside the event horizon of BH is
called effective potential. However, Hawking radiations cannot cross
the barrier completely, therefore, the observer cannot observe the
same spectrum of these radiations. Consequently, an observer outside
the event horizon will observe the emission rate differently from
the real one. Thus, the emission rate for the observer is expressed
as
\begin{equation}\nonumber
\gamma(w)=\left(\frac{|G_{M,l}|^2d^3k}{(e^{\frac{w}{t_h}}\pm1)8
\pi^3}\right),
\end{equation}
where $|G_{M,l}|^{2}$ is a greybody factor (GF) based on the
frequency. The GF is the probability of waves arriving from infinity
and absorbed by BH \cite{9}-\cite{12}. This factor is more
significant as it changes the Hawking radiation formula and is used
to calculate the absorption cross-section of BH. It is observed that
Hawking temperature and entropy vary with respect to the size of BHs
\cite{13}-\cite{15}.

Ida et al \cite{16} used the scalar field in higher dimensions to
examine the GF of rotating BH in a low-frequency expansion. Creek et
al \cite{17} investigated the GF for Bardeen BH and checked the
emission rate of the scalar field by analytical and numerical
methods. Chen et al \cite{18} calculated the GF for d-dimensional BH
by using quintessence field and obtained that frequency increases as
the luminosity of radiation decreases. They also found that the
corresponding solutions reduce to the d-dimensional
Reissner-Nordstrom BH for the specific value of frequency. Crispino
et al \cite{19} investigated the absorption process of Schwarzschild
BH for non-minimally coupled scalar fields. Kanti et al \cite{20}
derived the GF for the scalar field by using higher-dimensional
Schwarzschild-de Sitter spacetime.

Jorge et al \cite{21} computed the GF for higher-dimensional
rotating BHs with the cosmological constant in a low-frequency
regime. Toshmatov et al \cite{22} measured the effect of charge as
well as absorption rate for regular BHs and found that charge
reduces the transmission factor for incident waves. Ahmad and
Saifullah \cite{23} used cylindrically symmetric spacetime and found
GF for the uncharged and massless scalar field. Dey and Chakrabarti
\cite{24} considered Bardeen-de Sitter spacetime and measured the
probability of absorption as well as quasinormal modes. Sharif and
Ama-Tul-Mughani \cite{25} examined the GF for rotating Bardeen and
Kerr-Newman BHs surrounded by quintessence. In a recent paper
\cite{26}, Sharif and Shaukat calculated the GF for a rotating
Bardeen BH surrounded by perfect fluid dark matter.

Born and Infeld \cite{27} used non-linear electrodynamics (NLED) to
ensure that the self-energy of a point-like charge is finite. Beato
and Garcia \cite{28} coupled general relativity with NLED to find a
non-singular BH solution. Cai et al \cite{29} coupled BHs with
Born-Infeld NLED and found the importance of the cosmological
constant in the stable region of BH. The static spherically
symmetric BH with NLED, minimally/non-minimally coupled with gravity
has become a subject of great interest for the researchers. Bolokhov
et al \cite{30} developed examples of minimally coupled BHs with
gravity having four different horizons. The non-minimally coupled
BHs with gravity are important for dark energy and inflation. Many
singular as well as non-singular BHs with NLED have been constructed
in \cite{31}.  No hair conjecture states that BH is described
completely by only three parameters like mass, charge and spin.
Chowdhury and Banerjee \cite{32} evaluated GF for Reissner-Nordstrom
BH endowed with a scalar hair and gave the counter-example of no
hair conjecture. They observed the opposite behavior of hairy scalar
as compared to charge parameter.

This paper explores the effective potential and GF for static
spherically symmetric BH with NLED. The paper is planned as follows.
In section \textbf{2}, we find the effective potential by using the
radial equation of motion and tortoise coordinate. We find solutions
of the radial equation near and far away from the horizons in
section \textbf{3}. Section \textbf{4} matches the obtained
solutions in the intermediate regime and finds GF. The summary of
the results is given in section \textbf{5}.

\section{Effective Potential}

In this section, we formulate the effective potential required for
Hawking radiations. We consider a static spherically symmetric BH
with NLED as \cite{33}
\begin{equation}\label{1}
ds^2=-h(r)dt^{2}+\frac{dr^2}{h(r)}+r^2(d\theta^{2}+\sin^2\theta
d\phi^2),
\end{equation}
where
\begin{equation}\label{2}
h(r)=1-\frac{2M}{r}+\frac{Q^2}{r^2}-\frac{r^2\alpha^2}{3}+2\alpha Q.
\end{equation}
Here, $\alpha$ is a coupling constant while $Q$ and $M$ represent
charge and mass of the BH, respectively. It is mentioned here that
Schwarzschild BH is recovered for $Q=\alpha=0$. To find the roots of
Eq.(\ref{2}), we take $h(r)=0$, which yields
\begin{equation}\label{3}
1-\frac{2M}{r}+\frac{Q^2}{r^2}-\frac{r^2\alpha^2}{3}+2\alpha Q=0.
\end{equation}
Its solution gives two horizons as event horizon (near to BH) and
cosmological horizon (far away from BH). In order to analyze the
propagation of scalar field, we assume that particles and gravity
are minimally coupled and use the Klein-Gorden equation as
\begin{equation}\label{4}
\partial_{\delta}[\sqrt{-g}g^{\delta\nu}\partial_{\nu}\Sigma]=0.
\end{equation}
This equation is solved by using separation of variables method as
\begin{equation}\nonumber
\Sigma=\exp(-\iota wt)R_{wlm}(r)Y^{l}_{m}(\theta,\phi),
\end{equation}
where $Y^{l}_{m}(\theta,\phi)$ shows angular behavior. Thus,
Eq.(\ref{4}) turns out to be
\begin{eqnarray}\label{5}
&&\frac{1}{r^2}\left(r^2h\frac{d}{dr}R_{wlm}\right)_{,r}+\left(\frac
{w^2}{h}-\frac{\lambda_{l}}{r^2}\right)R_{wlm}=0,\\\label{6}&&\frac
{1}{\sin\theta}\left(\sin\theta\frac{\partial Y_{m}^{l}}{\partial
\theta}\right)_{,\theta}+\frac{1}{\sin^2\theta}\frac{(\partial^2Y^{l}
_{m})}{\partial\phi^2}+\lambda_lY_{m}^{l}=0,
\end{eqnarray}
where $\lambda_{l}$ is a separation constant. This determines the
connection between radial and angular equations \cite{34}. The power
series of separation constant is given by \cite{35}
\begin{eqnarray}\label{7}
\lambda_{l}=\sum_{k=0}^{\infty}(aw)^{k}F^{lm}_{k},\quad\lambda_{l}=l
(l+1)+O(s,w).
\end{eqnarray}
Here, orbital angular momentum $(l)$ satisfies the conditions
$l\geq|{m}|$ and $\frac{l-|m|}{2}\in({0,\mathbb{Z}})$, $\mathbb{Z}$
represents the set of integers. First, we find the potential which
affects the outcoming radiation from the BH. For this purpose, we
redefine the radial equation and introduce a new coordinate called
tortoise coordinate $(v_{\star})$ as
\begin{eqnarray}\nonumber
R_{wlm}(r)=\frac{T_{wlm}(r)}{r},\quad\frac{dv_{\star}}{dr}=\frac{1}{h},
\quad\frac{d}{dv_{\star}}=h\frac{d}{dr},\quad\frac{d^2}{dv_{\star}^2}=h
\left(\frac{d^2}{dr^2}+\frac{dh}{dr}\frac{d}{dr}\right).
\end{eqnarray}
We can observe that $r\rightarrow r_h \Rightarrow
v_{\star}\rightarrow -\infty$ and $r \rightarrow \infty$
$\Rightarrow v_\star\rightarrow \infty$. The radial equation works
for both (inside/outside) event horizons. The corresponding radial
equation becomes
\begin{equation}\label{8}
(\frac{d^{2}}{dv_\star^{2}}-V_{ef})T_{wlm}=0,
\end{equation}
where
$V_{ef}=h(\frac{1}{r}\frac{dh}{dr}-w^2+\frac{\lambda_{l}}{r^2})$ is
the effective potential which vanishes at $h=0$. We can analyze its
behavior through graphs for different physical parameters.
\begin{figure}
\epsfig{file=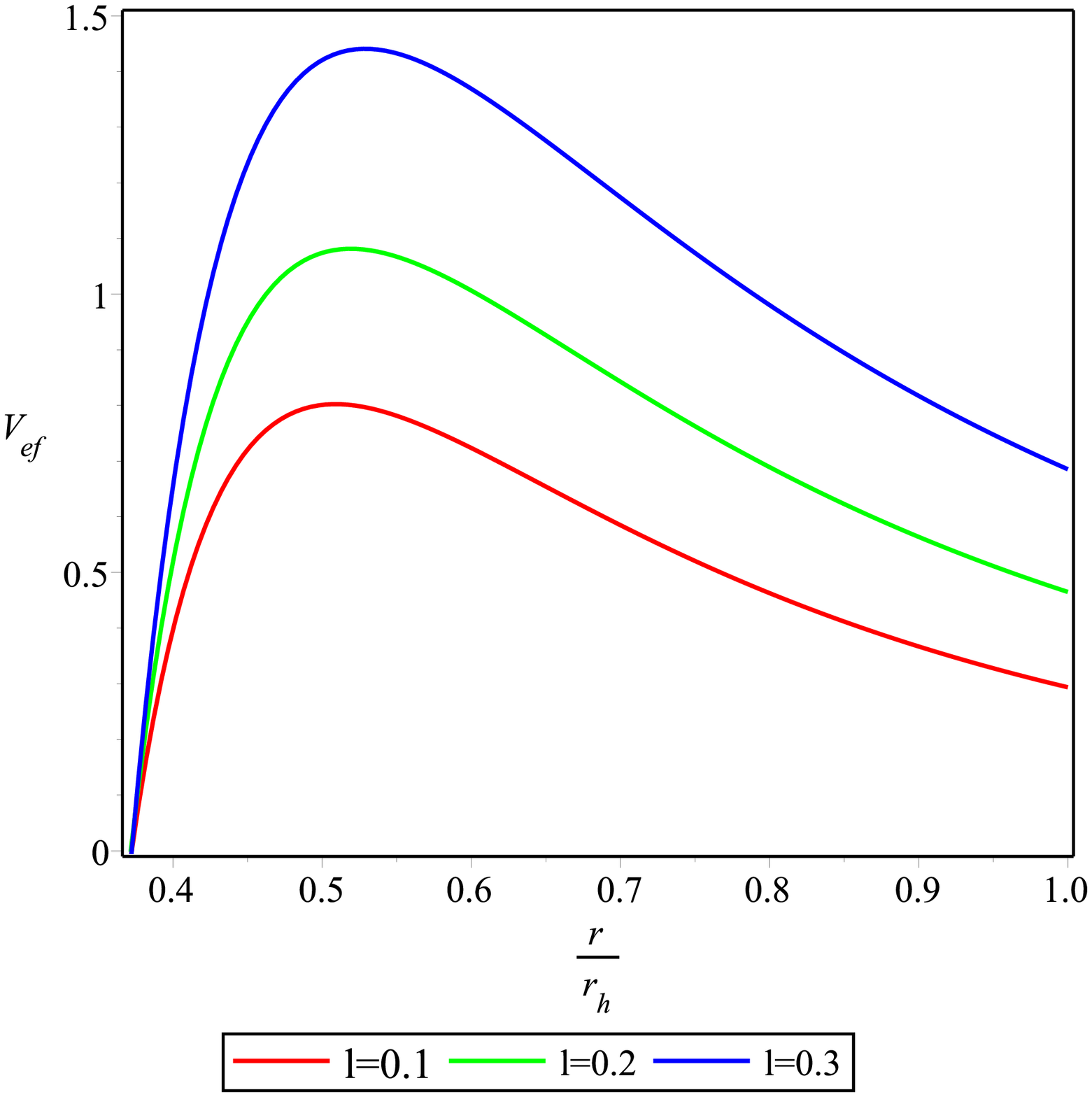,width=0.5\linewidth}
\epsfig{file=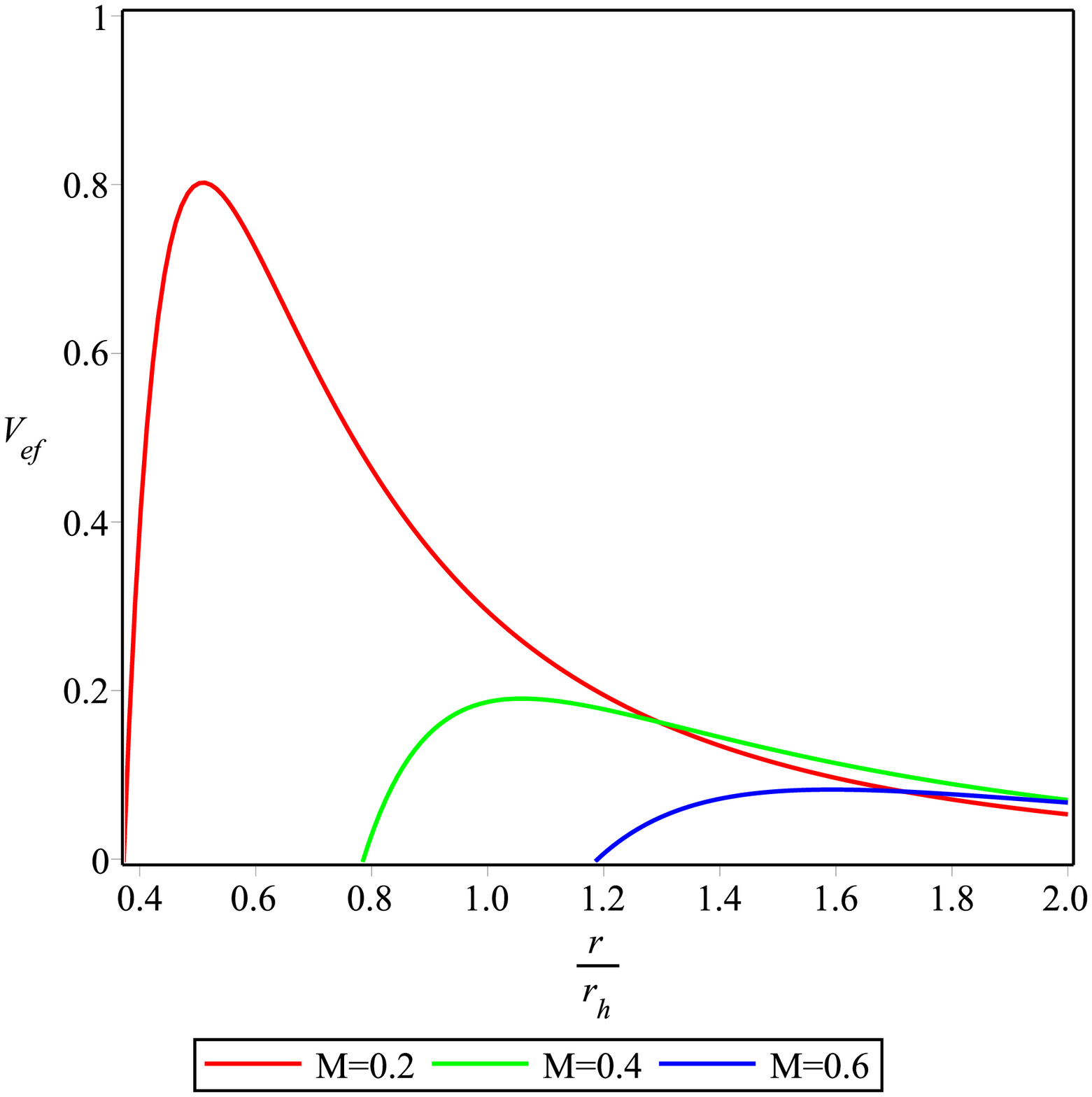,width=0.5\linewidth}\caption{Plots of effective
potential versus $\frac{r}{r_{h}}$ with $M=0.2$ (left) and with
$l=0.1$ (right) for $Q=0.1$, $\alpha=0.01$ and $w=0.1$.}
\end{figure}
\begin{figure}
\epsfig{file=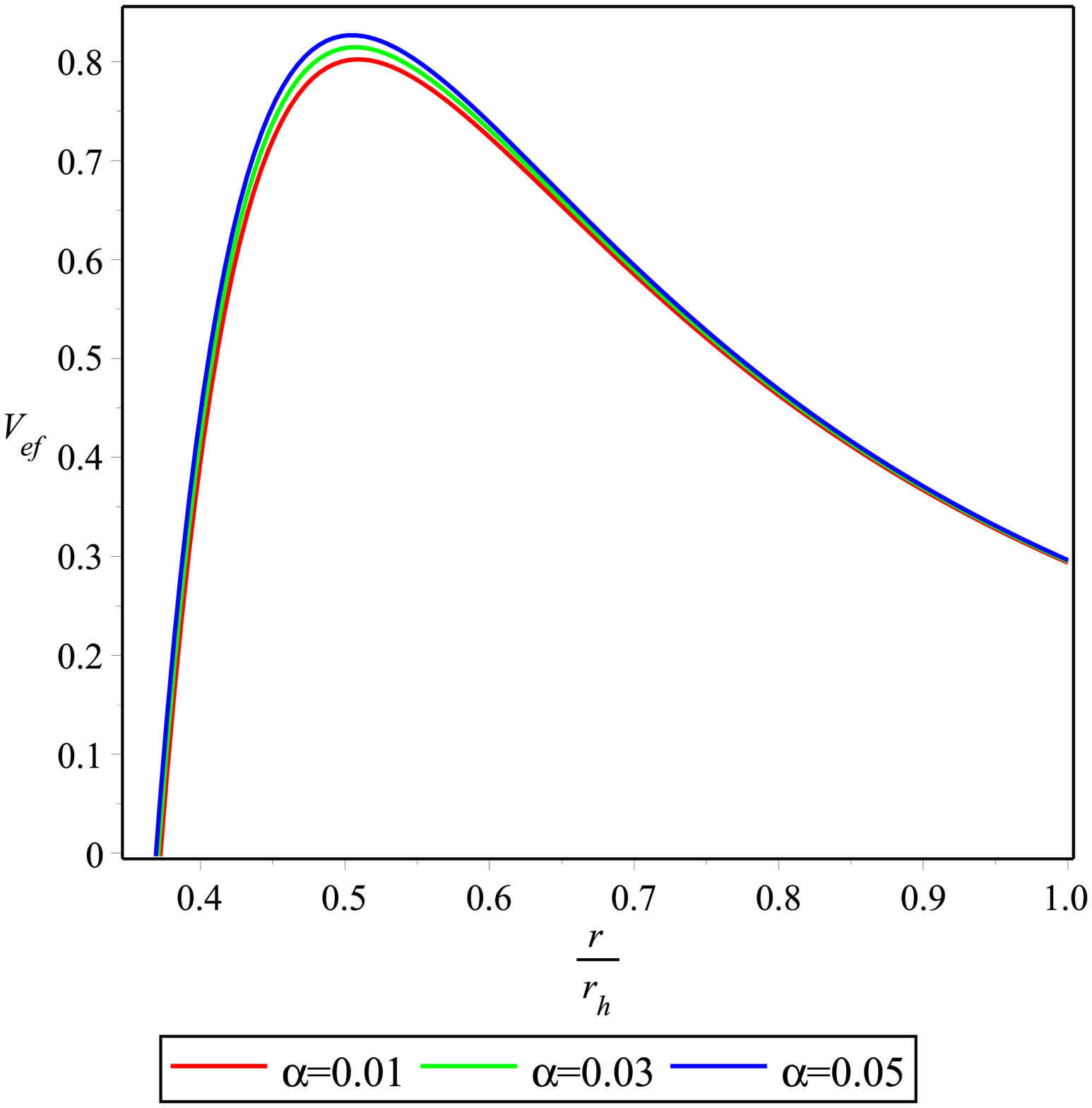,width=0.5\linewidth}
\epsfig{file=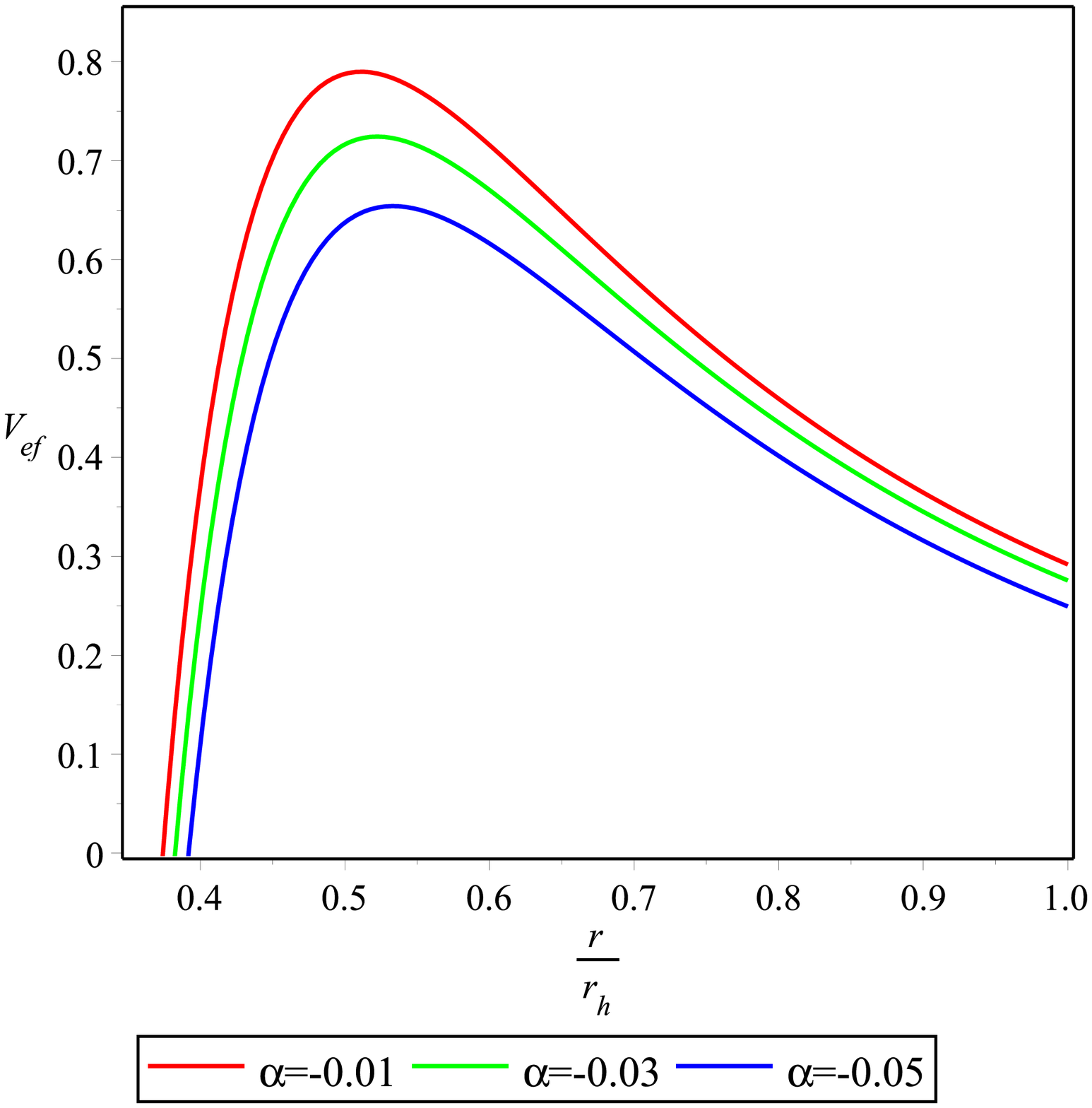,width=0.5\linewidth}\caption{Plots of effective
potential versus $\frac{r}{r_{h}}$ with $\alpha>0$ (left) and
$\alpha<0$ (right) for $M=0.2$, $Q=0.1$ and $w=0.1$.}
\end{figure}

The behavior of effective potential for various values of mass and
angular momentum is given in Figure \textbf{1}. The graph in the
right panel shows that the height of effective potential is higher
for the smaller value of mass corresponding to the radial
coordinate. The angular momentum in the left graph has a direct
relation with the effective potential which minimizes the GF. In
Figure \textbf{2}, the behavior of the effective potential shows
direct relation with the coupling constant that lowers the emission
rate. Figure \textbf{3} (left plot) shows that the effective
potential decreases with the increase of charge parameter. This
behavior shows that the presence of charge parameter decreases the
effective potential which increases the absorption probability. The
right graph shows that the effective potential decreases with the
frequency of Hawking radiations which indicates the increasing
behavior of the GF.
\begin{figure}
\epsfig{file=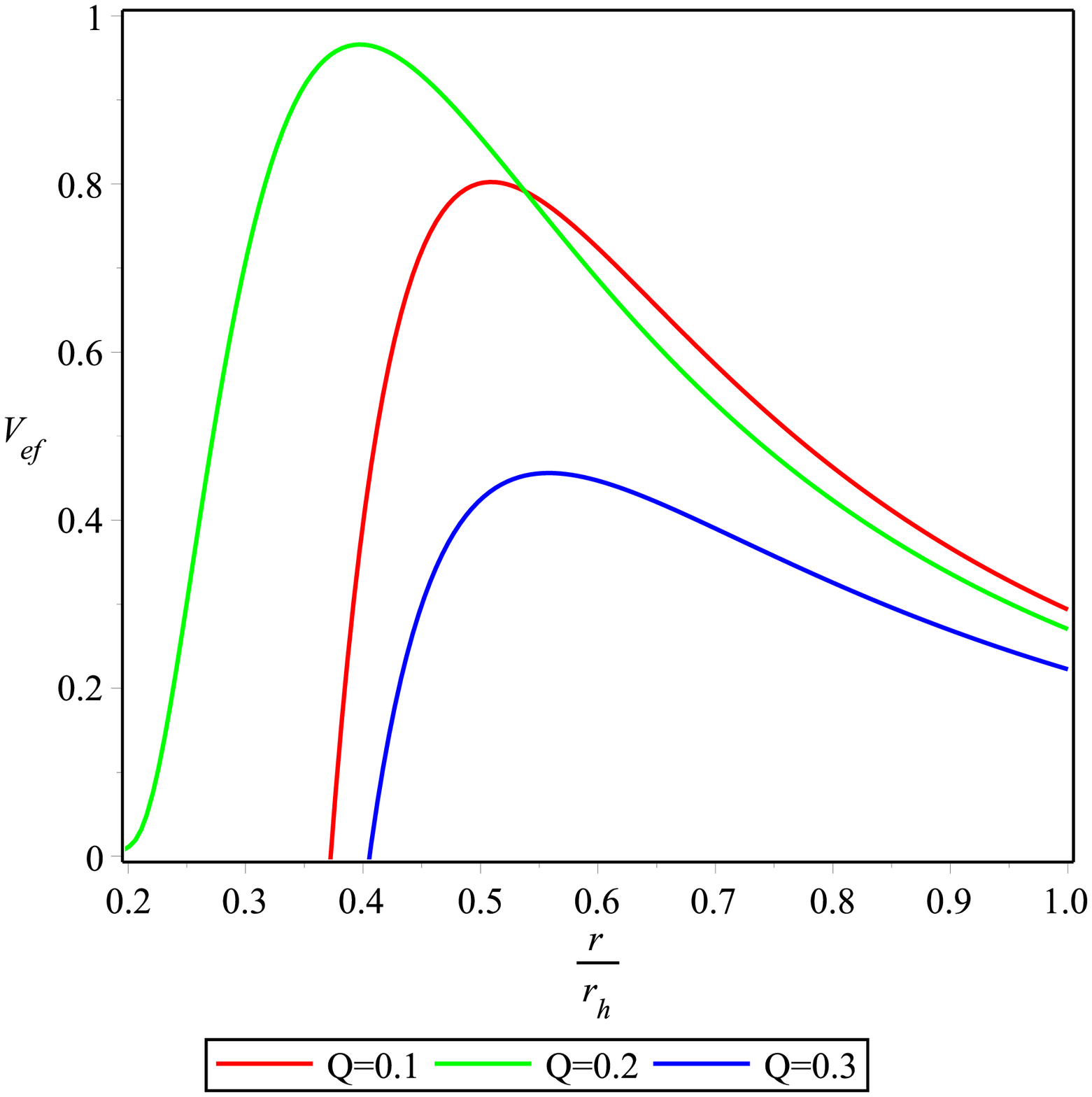,width=0.5\linewidth}
\epsfig{file=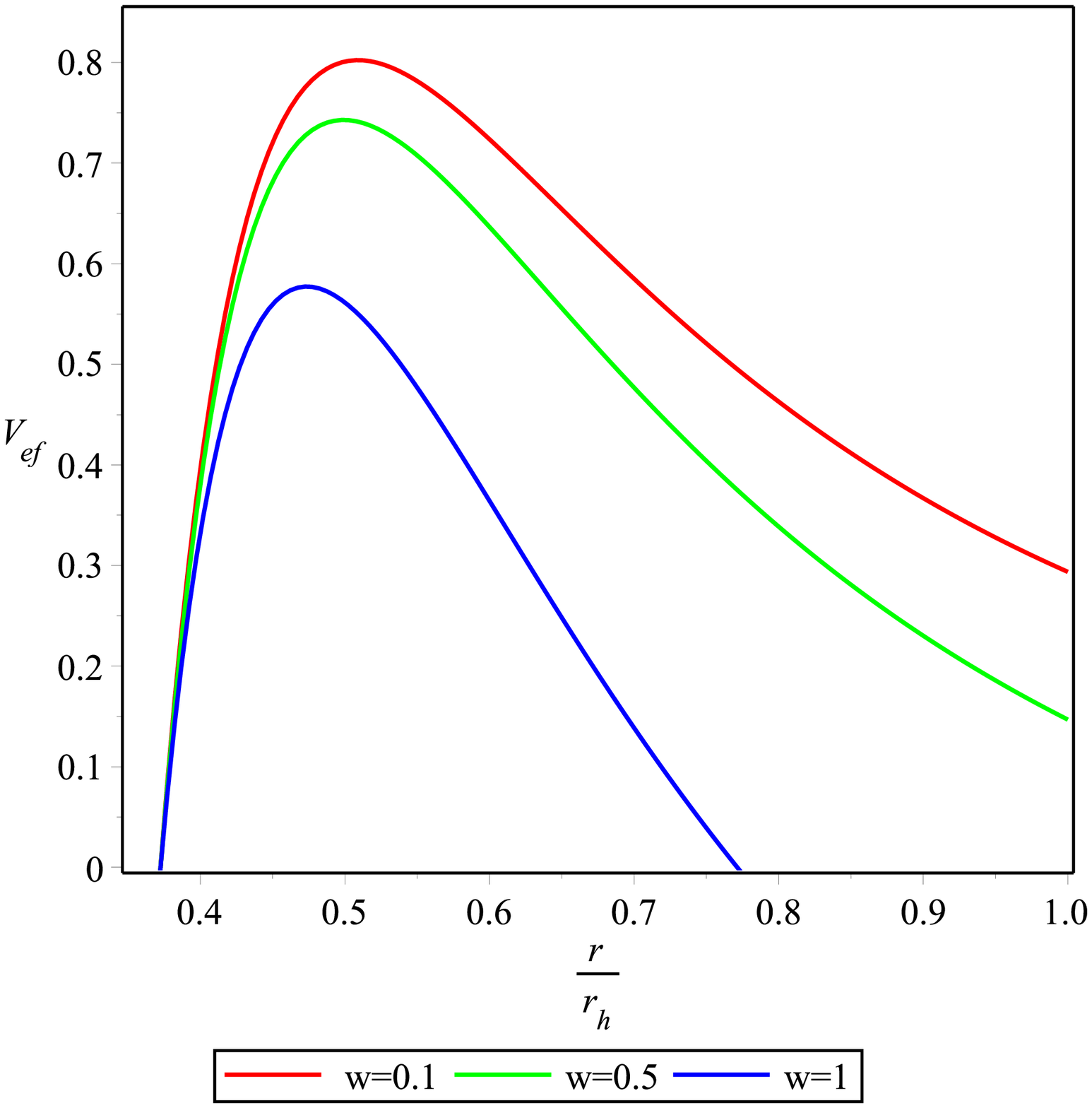,width=0.5\linewidth}\caption{Plots of effective
potential versus $\frac{r}{r_{h}}$ with $w=0.1$ (left) and $Q=0.1$
(right) for $l=0.1$, $M=0.2$, $l=0.1$ and $\alpha=0.01$.}
\end{figure}

\section{Greybody Factor}

This section evaluates the GF by analytical approach. We use radial
equation and apply transformations near and far away from the event
horizon. We then match both solutions in the middle region. These
transformations give suitable equations which can be solved
analytically. The first transformation near the horizon is
\begin{eqnarray}\label{9}
r\rightarrow
\Theta=\frac{1-\frac{2M}{r}+\frac{Q^2}{r^2}-\frac{r^{2}\alpha^2}{3}+2\alpha
Q}{{1-\frac{r^2\alpha^2}{3}}},\quad\frac{d\Theta}{dr}=\frac{(1-\Theta)P}
{r(3-r^{2}\alpha^{2})},
\end{eqnarray}
where
\begin{equation}\nonumber
P=\frac{2(3Mr+3Q^2-(3Mr+2Q^2)\alpha^2r^2-2\alpha^3Qr^{4})}{2Mr-Q^2
-2Q\alpha r^{2}}.
\end{equation}
Using Eq.(\ref{9}) in the radial equation, we have
\begin{equation}\label{10}
\Theta(1-\Theta)\frac{d^{2}R_{wlm}}{d\Theta^{2}}+(B-A\Theta)\frac{dR_{wlm}}
{d\Theta}+\frac{1}{(1-\Theta)P^2}(\frac{\zeta_{h}}{\Theta}-\xi_{h})R_{wlm}=0,
\end{equation}
where
\begin{eqnarray}\nonumber
A=\frac{3(rh(1-\Theta)P)'}{(1-\Theta)P^{2}},\quad
B=-\frac{2r^{2}\alpha^{2}}{P},~\zeta_{h}=9w^{2}r^{2},
~\xi_{h}=3(l^2-l)(3-r^{2}\alpha^{2}).
\end{eqnarray}
We use the transformation
$(R_{wlm}(\Theta)=\Theta^{\pi_{1}}(1-\Theta)^{\tau_{1}}W_{wlm}(\Theta))$
in Eq.(\ref{10}) to obtain hypergeometric (HG) equation. The
corresponding equation becomes
\begin{eqnarray}\nonumber
&&\Theta(1-\Theta)\frac{d^2W_{wlm}}{d\Theta^2}+[2\pi_{1}+B-(2\pi_1+2
\tau_1+A)\Theta]\frac{dW_{wlm}}{d\Theta}+\bigg[\frac{1}{\Theta}(\pi_{1}
^{2}-\pi_1\\\nonumber&&+B\pi_1+\frac{\zeta_h}{P^2})+\frac{1}{1-\Theta}
(\tau_{1}^{2}-\tau_1-B\tau_1+A\tau_1+\frac{\zeta_h}{P^2}-\frac{\xi_h}
{P^2})-(\tau_1+\pi_1)A\\\nonumber&&-\pi_{1}^{2}-2\pi_1\tau_1+\pi_1-
\tau_{1}^{2}+\tau_1\bigg]W_{wlm}=0.
\end{eqnarray}
In order to find the power coefficients, we assume that the
coefficients of $\frac{1}{\Theta}$ and $\frac{1}{1-\Theta}$ are zero
\begin{eqnarray}\nonumber
\pi_{1}^{2}-\pi_{1}(1-B)+\frac{\zeta_{h}}{P^{2}}=0,\quad\tau_{1}^{2}
-\tau_1(1-A-B)+\frac{\zeta_{h}}{P^{2}}-\frac{\xi_{h}}{P^2}=0.
\end{eqnarray}
The corresponding radial equation (\ref{5}) turns out to be
\begin{equation}\label{11}
\Theta(1-\Theta)\frac{d^2W_{wlm}}{d\Theta^2}+(k_{1}-(i_{1}+j_{1}+1)
\Theta)\frac{dW_{wlm}}{d\Theta}-i_{1}j_{1}W_{wlm}=0,
\end{equation}
where $i_{1}=\pi_{1}+\tau_{1}$, $j_{1}=\pi_{1}+\tau_{1}+A-1$,
$k_{1}=2\pi_{1}+B$. The general solution of this equation for near
the horizon is
\begin{eqnarray}\nonumber
(R_{wlm})_{nh}(\Theta)&=&I_{1}\Theta^{\pi_{1}}(1-\Theta)^{\tau_{1}}
F(i_{1},j_{1},k_{1};\Theta)+I_{2}\Theta^{-\pi_{1}}(1-\Theta)
^{\tau_{1}}\\\nonumber&\times&F(1-k_{1}+i_{1},1-k_{1}+j_{1},2-k_{1}
;\Theta),
\end{eqnarray}
where $I_1$ and $I_2$ are constants and
\begin{eqnarray}\nonumber
\pi_{1}^{\pm}&=&\frac{1}{2}\left[(1-B)\pm\sqrt{(1-B^2)-4\frac{\zeta_h}
{P^2}}\right],\\\nonumber\tau_{1}^{\pm}&=&\frac{1}{2}\left[(1+B-A)\pm
\sqrt{(1-A+B)^2+4(\frac{\xi_h}{P^2}\frac{\zeta_h}{P^2})}\right].
\end{eqnarray}

Now, we apply the boundary conditions as outgoing wave is not
observed near the event horizon. We can choose $I_1=0$, or $I_2=0$
depending upon whether $\pi_1$ is positive or negative. It is found
that the choice of $\pi_1$ does not affect the choice of constants,
therefore, we take $\pi_{1}=\pi_{1}^{-}$ with $I_2=0$. The
corresponding solution becomes
\begin{equation}\label{12}
(R_{wlm})_{nh}(\Theta)=I_{1}\Theta^{\pi_{1}}(1-\Theta)^{\tau_{1}}
F(i_{1},j_{1},k_{1};\Theta).
\end{equation}
Applying the same procedure for the cosmological horizon, we have
\begin{eqnarray}\label{13}
\Upsilon(r)=\frac{h}{r^{2}}=\frac{1}{r^2}-\frac{\alpha^2}{3},\quad\frac{d\Upsilon}{dr}
=\frac{(1-\Upsilon)D}{r},
\end{eqnarray}
where $D(r)=\frac{-6}{r^2(3+\alpha^2)-3}$. Using the transformation
\begin{equation}\nonumber
R_{wlm}(\Upsilon)=\Upsilon^{\pi_{2}}(1-\Upsilon)^{\tau_{2}}W_{wlm}(\Upsilon),
\end{equation}
and Eq.(\ref{13}) in (\ref{5}), we obtain
\begin{eqnarray}\nonumber
&&\Upsilon(1-\Upsilon)\frac{d^2W_{wlm}}{d\Upsilon^2}+(2\pi_{2}+B_\star
-(2\pi_2+2\tau_2+A_\star)\Upsilon)\frac{dW_{wlm}}{d\Upsilon}\\\nonumber&&
+\bigg[(\pi_{2}^{2}-\pi_2+B_{\star}\pi_2+\frac{\zeta_\star}{D^2})\frac{1}
{\Upsilon}+(\tau_{2}^{2}-\tau_2-B_{\star}\tau_2+A_{\star}\tau_2+\frac{\zeta
_\star}{D^2}-\frac{\xi_\star}{D^2})\\\nonumber&&\times\frac{1}{1-\Upsilon}
-\pi_{2}^{2}-2\pi_2\tau_2-A_{\star}(\pi_2-\tau_2+\pi_2+\tau_2)\bigg]W_{wlm}=0.
\end{eqnarray}
To find the power coefficients $\pi_2$ and $\tau_2$, we assume the
coefficients of $\frac{1}{\Upsilon}$ and $\frac{1}{1-\Upsilon}$ to
be zero
\begin{eqnarray}\nonumber
\pi_{2}^{2}-\pi_{2}(1-B_\star)+\frac{\zeta_{\star}}{D^{2}}=0,\quad\tau_{2}^{2}
-\tau_{2}(1-A_{\star}-B_{\star})+\frac{\zeta_{\star}}{D^{2}}-\frac{\xi_{\star}}
{D^2}=0.
\end{eqnarray}
The corresponding radial equation in the form of HG equation turns
out to be
\begin{equation}\label{14}
\Upsilon(1-\Upsilon)\frac{d^2W_{wlm}}{d\Upsilon^2}+(k_{2}-(i_{2}+j_{2}+1)\Upsilon)
\frac{dW_{wlm}}{d\Upsilon}-i_{2}j_{2}W_{wlm}=0,
\end{equation}
where $i_{2}=\pi_{2}+\tau_{2}, j_{2}=\pi_{2}+\tau_{2}+A_{\star}-1,
k_{2}=2\pi_{2}+B_{\star}$. The general solution of HG equation is
\begin{eqnarray}\nonumber
(R_{wlm})_{fh}(\Upsilon)&=&J_{1}\Upsilon^{\pi_{2}}(1-\Upsilon)^{\tau_{2}}F(i_{2},j_{2}
,k_{2};\Upsilon)+J_{2}\Upsilon^{\pi_{2}}(1-\Upsilon)^{\tau_{2}}\\\label{15}
&\times&F(1+i_{2}-k_{2},1+j_{2}-k_{2},2-k_{2};\Upsilon),
\end{eqnarray}
where $J_1$ and $J_2$ are arbitrary constants.

\section{Matching Regime}

Here, we match the obtained solutions at the event and cosmological
horizons in the intermediate region corresponding to $r$. For this
reason, we stretch the event horizon towards the cosmological
horizon by replacing the argument $\Theta$ by $1-\Theta$ of HG
function in Eq.(\ref{12}) and obtain
\begin{eqnarray}\nonumber
(R_{wlm})_{nh}(\Theta)&=&I_{1}\Theta^{\pi_{1}}(1-\Theta)^{\tau_{1}}\bigg[\frac
{\Gamma(-i_{1}-j_{1}+k_{1})\Gamma(k_{1})}{\Gamma(k_{1}-i_{1})\Gamma(k_{1}-j_{1}
)}F(i_{1},j_{1},k_{1};1-\Theta)\\\nonumber&+&(1-\Theta)^{-i_1-j_1+k_1}\frac{
\Gamma(k_1)\Gamma(i_1+j_1-k_1)}{\Gamma(j_1)\Gamma(i_1)}\\\nonumber&\times&
F(k_1-i_1,k_1-j_1,1-i_1-j_1+k_1;1-\Theta)\bigg].
\end{eqnarray}
Using Eqs.(\ref{3}) and (\ref{9}), we have
\begin{equation}\nonumber
1-\Theta=\frac{3(2Mr-Q^2-2Qr^2\alpha)}{r^2(3-r^2\alpha^2)}.
\end{equation}
The extended event horizon for $\Theta\rightarrow{1}$ and $r\gg r_h$
is
\begin{eqnarray}\nonumber
(1-\Theta)^{\tau_1}\simeq (Q_{\star}^{2}+2 \alpha
Q)^{\tau_1}(\frac{r}{r_h})^{\tau_1}\Rightarrow(1-\Theta)^{\tau_1}\simeq
(Q_{\star}^{2}+2\alpha Q)^{-l}(\frac{r}{r_h})^{-l},
\end{eqnarray}
and
\begin{eqnarray}\nonumber
(1-\Theta)^{\tau_1+k_1-i_1-j_1}&\simeq&(Q_{\star}^{2}+2\alpha
Q)^{-\tau_1+B-A+1}(\frac{r}{r_h})^{-\tau_1+B-A+1}
\\\nonumber&\simeq&
(Q_{\star}^{2}+2\alpha Q)^{l+1}(\frac{r}{r_h})^{l+1},
\end{eqnarray}
where $Q_{\star}=\frac{Q}{r_{h}}$. It is worthwhile to mention here
that the constraints are valid for small values of charge and
coupling parameters. In an intermediate zone, both parts of near
horizon BH solution can be written as
\begin{eqnarray}\nonumber
(1-\Theta)^{\tau_1}\simeq (Q_{\star}^{2}+2\alpha
Q)^{_l}(\frac{r}{r_h})^{-l}, \quad (1-\Theta)^{\tau_1}\simeq
(Q_{\star}^{2}+2\alpha Q)^{1+l}(\frac{r}{r_h})^{1+l}.
\end{eqnarray}
Finally, the solution on the event horizon is
\begin{equation}\label{16}
(R_{wlm})_{nh}(\Theta)=I'(\frac{r}{r_h})^{-l}+I'_{2}(\frac{r}{r_h})^{l+1},
\end{equation}
with
\begin{eqnarray}\nonumber
I'_1&=&I_1(Q_{\star}^{2}+2\alpha
Q)^{-l}\frac{\Gamma(-i_{1}-j_{1}+k_{1})\Gamma(k_{1})}{\Gamma(k_{1}-i_{1})
\Gamma(k_{1}-j_{1})},\\\nonumber I'_2&=&(Q_{\star}^{2}+2\alpha
Q)^{l+1}\frac{\Gamma(k_1)\Gamma(i_1+j_1-k_1)}{\Gamma(j_1)\Gamma(i_1)}.
\end{eqnarray}

Now, we shift the cosmological horizon to the event horizon.
Therefore, we replace the argument $\Upsilon$ by $1-\Upsilon$ of HG
function in Eq.(\ref{15}) and obtain
\begin{eqnarray}\nonumber
&&(R_{wlm})_{fh}(\Upsilon)=J_1\Upsilon^{\pi_{2}}(1-\Upsilon)^{\tau_{2}}
\bigg[\frac{\Gamma(-i_{2}-j_{1}+k_{1})\Gamma(k_{2})}{\Gamma(k_{2}-i_{2})
\Gamma(k_{2}-j_{2})}F(i_{2},j_{2},k_{2};1-\Upsilon)\\\nonumber&&+F(k_2-i_2
,k_2-j_2,1-i_2-j_2+k_2;1-\Upsilon)\frac{\Gamma(k_2)\Gamma(i_2+j_2-k_2)}
{\Gamma(j_2)\Gamma(i_2)}\\\nonumber&&\times(1-\Upsilon)^{-i_2-j_2+k_2}
\bigg]+J_{2}\Theta^{-\pi_{2}}(1-\Upsilon)^{\tau_{2}}\bigg[\frac{\Gamma
(-i_{2}-j_{2}+k_{2})\Gamma(2-k_{2})}{\Gamma(1-i_{2})\Gamma(1-j_{2}))}
\\\nonumber&&\times F(-k_2+i_2+1,j_2-k_2+1,2-k_2;1-\Upsilon)+(1-\Theta)
^{-i_2-j_2+k_2}\\\label{17}&&\times\frac{\Gamma(2-k_2)\Gamma(i_2+j_2-k_2)}
{\Gamma(1-i_2)\Gamma(1-j_2)}F(1-i_2,1-j_2,1-i_2-j_2+k_1;1-\Upsilon)\bigg].
\end{eqnarray}
To find the solution of cosmological horizon, we consider
$\Upsilon(r_f)\rightarrow 0$, so that the corresponding
Eq.(\ref{13}) becomes
\begin{equation}\label{18}
1-\Upsilon=\frac{r}{r_f}\left(\frac{1}{rr_f}-\frac{r_f}{r^3}+\frac{r_f}{r}\right).
\end{equation}
This equation can be written in the following form
\begin{eqnarray}\nonumber
(1-\Upsilon)^{\tau_2}\simeq\left(\frac{r}{r_{f}}\right)^{-l}\left(\frac{1}{rr_f}
-\frac{r_f}{r^3}+\frac{r_f}{r}\right)^{-l},
\end{eqnarray}
and
\begin{eqnarray}\nonumber
(1-\Upsilon)^{\tau_2-i_2-j_2
+k_2}\simeq\left(\frac{r}{r_{f}}\right)^{l+1}\left(\frac{1}{rr_f}-\frac{r_f}{r^3}
+\frac{r_f}{r}\right)^{l+1}.
\end{eqnarray}
The corresponding Eq.(\ref{17}) turns out to be
\begin{equation}\label{19}
R_{fh}=(H'_1J_1+H'_2J_2)\left(\frac{r}{r_f}\right)^{-l}+(H'_3J_1+H'_4J_2)\left(\frac{r}
{r_f}\right)^{l+1},
\end{equation}
where
\begin{eqnarray}\nonumber
&&H'_1=\frac{\Gamma(k_2)\Gamma(k_2-i_2-j_2)}{\Gamma(k_2-i_2)\Gamma(k_2-j_2)}
\left(\frac{1}{rr_f}-\frac{r_f}{r^3}+\frac{r_f}{r}\right)^{-l},\\\nonumber
&&H'_2=\frac{\Gamma(2-k_2)\Gamma(k_2-i_2-j_2)}{\Gamma(1-i_2)\Gamma(1-j_2)}
\left(\frac{1}{rr_f}-\frac{r_f}{r^3}+\frac{r_f}{r}\right)^{-l},\\\nonumber
&&H'_3=\frac{\Gamma(k_2)\Gamma(-k_2+i_2+j_2)}{\Gamma(i_2)\Gamma(j_2)}
\left(\frac{1}{rr_f}-\frac{r_f}{r^3}+\frac{r_f}{r}\right)^{l+1},
\\\nonumber&&H'_4=\frac{\Gamma(2-k_2)\Gamma(-k_2+i_2+j_2)}{\Gamma(1-i_2)
\Gamma(1-j_2)}\left(\frac{1}{rr_f}-\frac{r_f}{r^3}+\frac{r_f}{r}\right)^{l+1}.
\end{eqnarray}
Comparing both the solutions, we obtain
\begin{equation}\nonumber
I'_1=H'_1J_1+H'_2J_2,I'_2=H'_3J_1+H'_4J_2,
\end{equation}
where
\begin{eqnarray}\label{20}
J_1=\frac{I'_1H'_4-I'_2H'_2}{H'_1H'_4-H'_2H'_3},\quad
J_2=\frac{I'_1H'_3-I'_2H'_1}{H'_2H'_3-H'_1H'_4}.
\end{eqnarray}
Consequently, the expression for GF is given as
\begin{equation}\label{21}
|G_{M,l}|^{2}=1-\left|\frac{J_2}{J_1}\right|^{2},
\end{equation}
which takes the following form using Eq.(\ref{20})
\begin{equation}\nonumber
|G_{M,l}|^{2}=1-\left|\frac{I'_1H'_3-I'_2H'_1}
{I'_1H'_4-I'_2H'_2}\right|^{2}.
\end{equation}
This is the final expression of GF for the static spherically
symmetric BH with NLED. It is observed that waves pass through the
cosmological horizon which is far away from the event horizon. Then
the waves either reflect or move forward implying that there is a
connection between the frequency and the effective potential. It is
noted that the waves must be of high frequency to cross the barrier
easily.
\begin{figure}
\epsfig{file=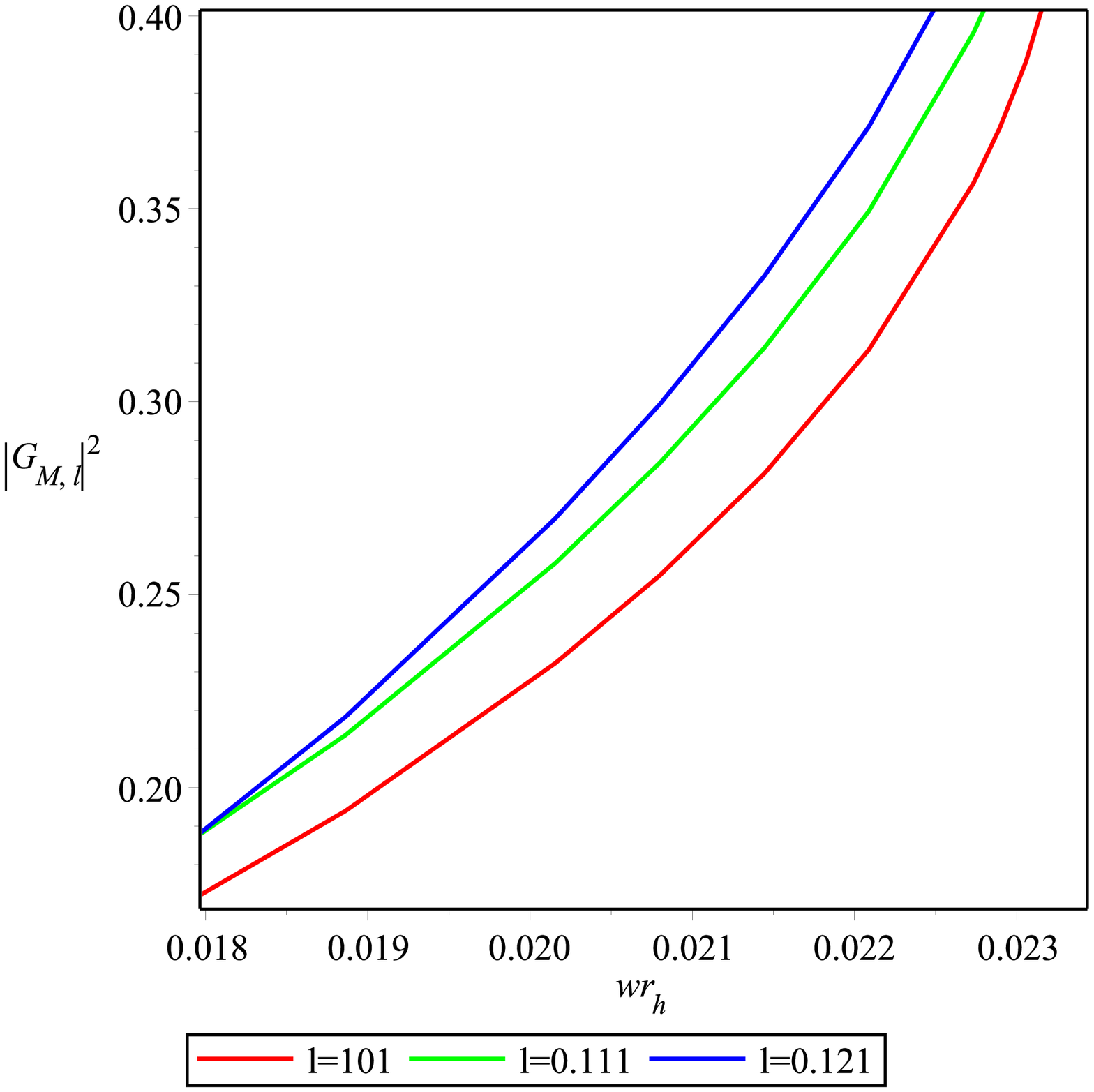,width=0.5\linewidth}
\epsfig{file=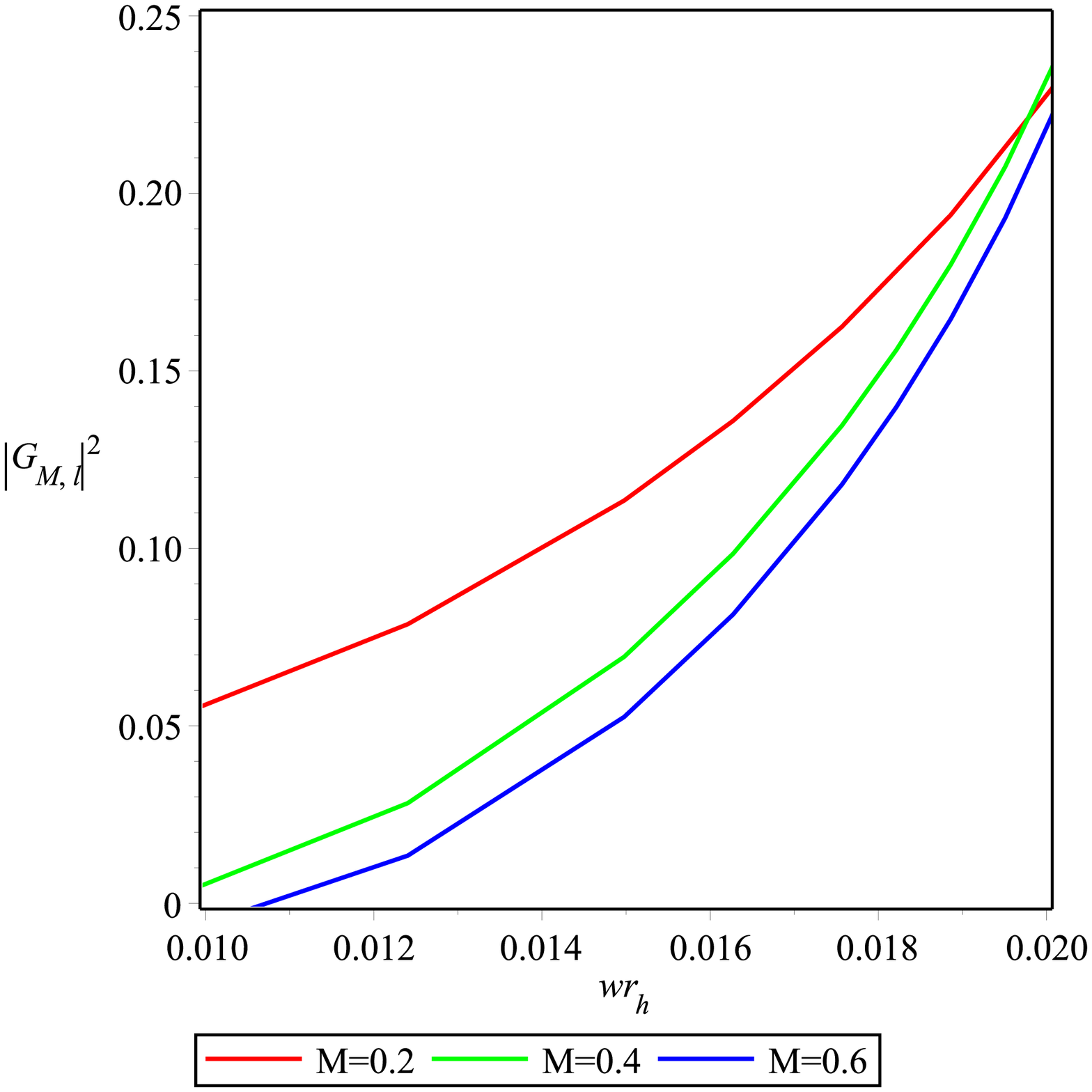,width=0.5\linewidth}\caption{Plots of GF versus
$wr_{h}$ with $M=0.2$ (left) and with $l=0.101$ (right) for
$r=10.5$, $Q=0.101$ and $\alpha=0.05$.}
\end{figure}
\begin{figure}
\epsfig{file=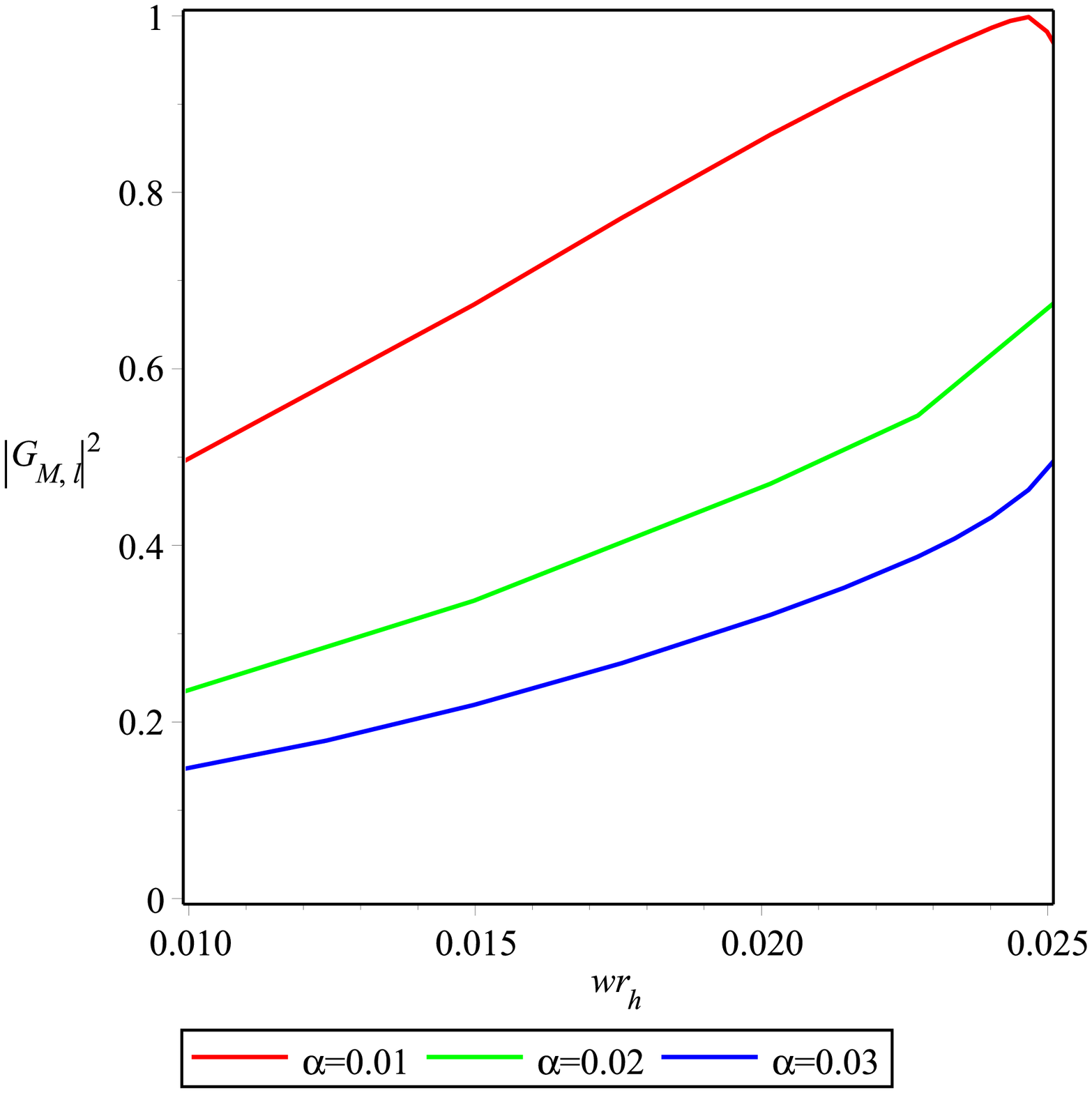,width=0.5\linewidth}
\epsfig{file=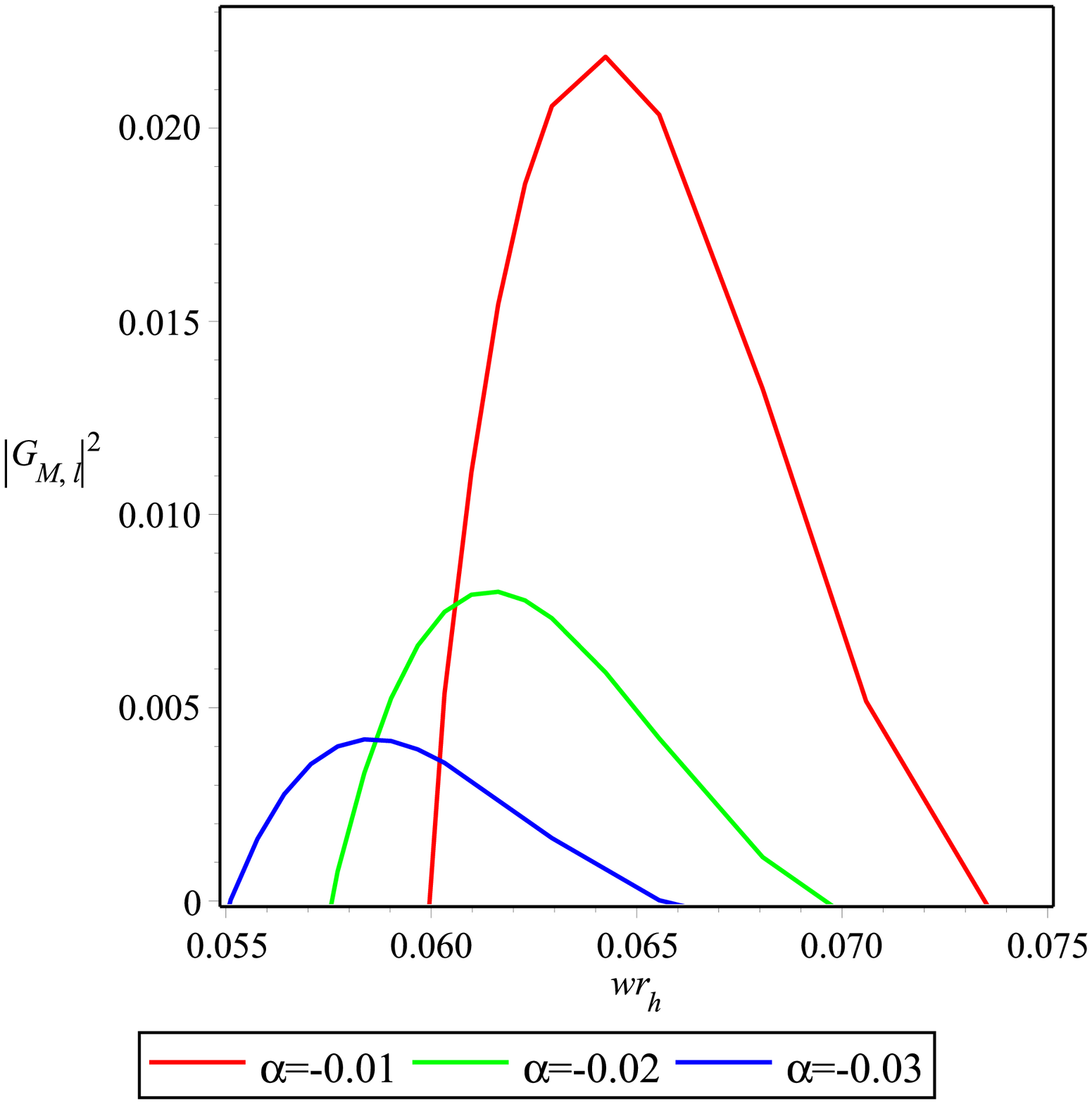,width=0.5\linewidth}\caption{Plots of GF versus
$wr_{h}$ with $\alpha>0$ (left) and with $\alpha<0$ (right) for
$M=0.2$, $l=0.101$, $r=10.5$ and $Q=0.101$.}
\end{figure}

We examine the effect of different physical parameters on GF through
graphs. Figure \textbf{4} shows the relationship between mass and
angular momentum, i.e., GF is smaller for higher values of mass and
larger for greater values of the angular momentum. This indicates
that GF decreases with the increase in the value of mass, and BH
with a larger value of angular momentum has a higher emission rate.
Figures \textbf{5}-\textbf{6} indicate that GF decreases with the
increase in the coupling and charge parameters. It shows that the
presence of the coupling and charge parameters decreases the
probability of absorption of the radiations. The GF has an inverse
relation with radius and disappears after specific values of the
radial coordinate. This indicates that BH with a larger size has a
lower emission rate and the GF.
\begin{figure}
\epsfig{file=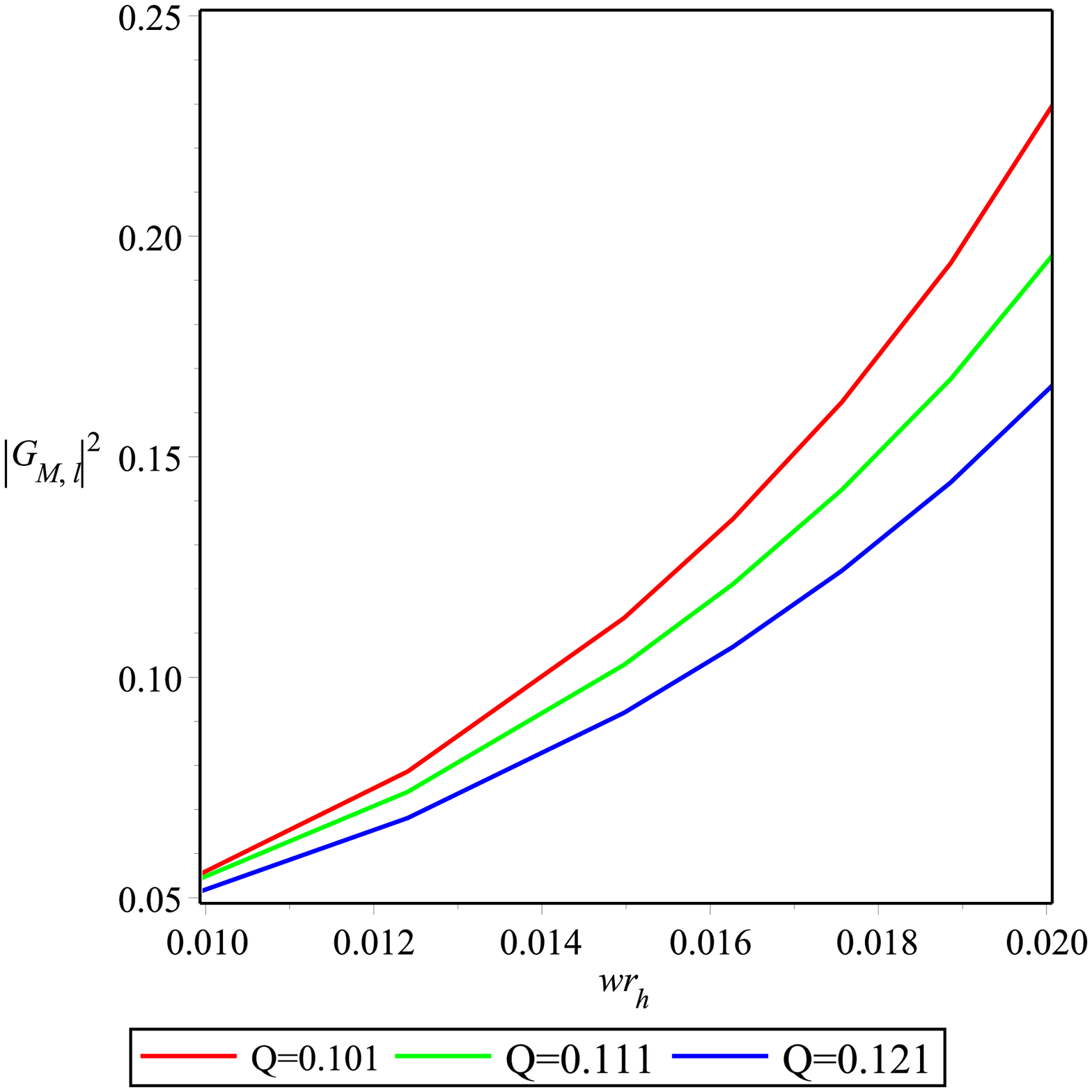,width=0.5\linewidth}
\epsfig{file=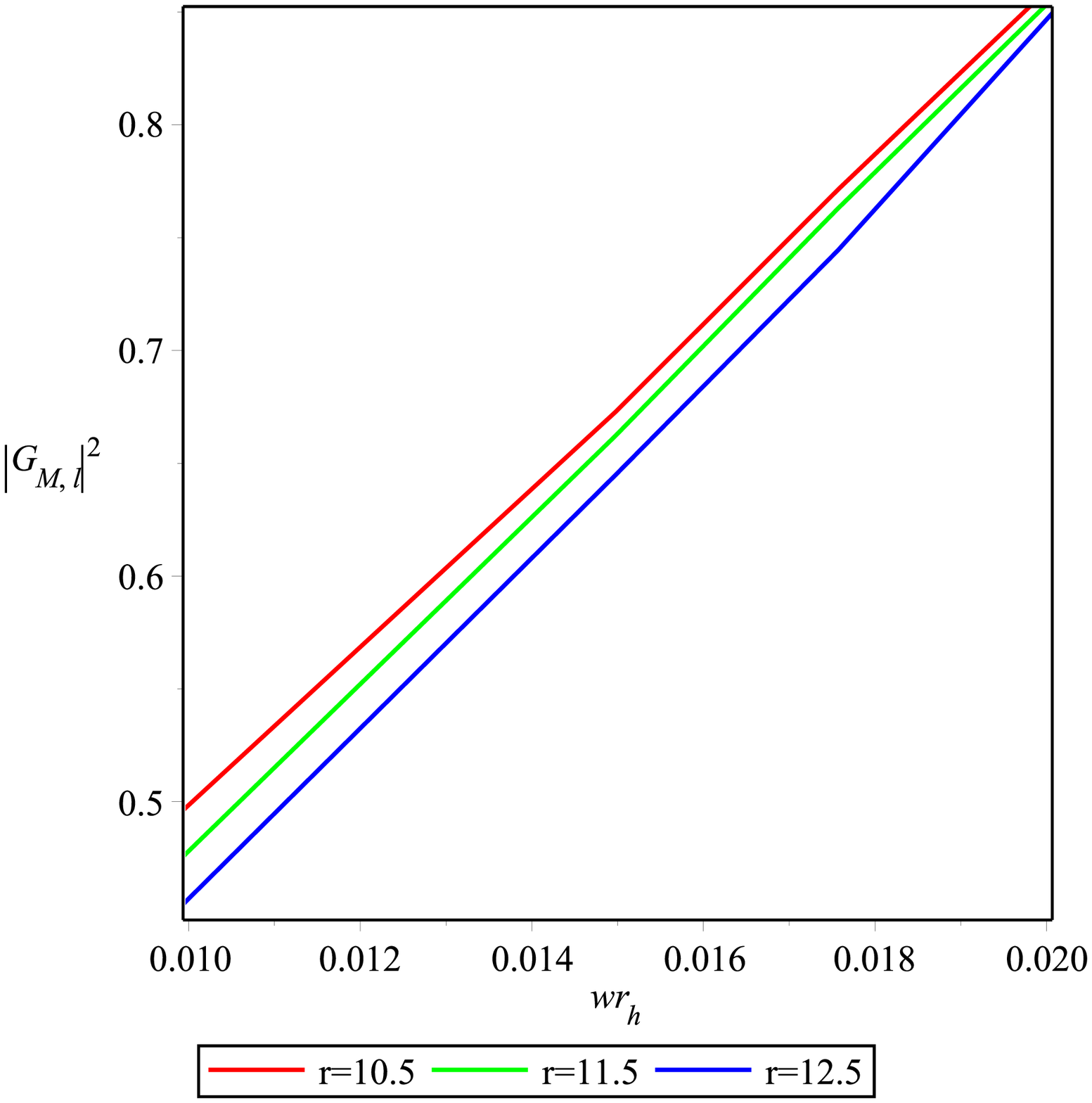,width=0.5\linewidth}\caption{Plots of GF versus
$wr_{h}$ with $r=10.5$ (left) and with $Q=0.101$ (right) for
$M=0.2$, $l=0.101$, and $\alpha=0.05$.}
\end{figure}

\section{Conclusions}

In this paper, we have analyzed the GF for static spherically
symmetric BH with NLED. We have used the Klein-Gorden equation of
motion and applied the separation of variables method to obtain the
radial and angular equations. We have then used tortoise coordinate
and transformed the radial equation into Schrodinger wave equation.
Further, we have obtained the effective potential for the absorption
of Hawking radiation and checked its behavior corresponding to the
coupling, charge and frequency parameters graphically.

We have worked near the BH and cosmological horizons and transformed
the radial equation into HG differential equations in both regions
to obtain the solutions. In the intermediate regime, we have matched
these two solutions by stretching the event horizon, compressing the
cosmological horizon and have found the expression for GF. The main
results of this paper are given as follows.
\begin{itemize}
\item We have found that the height of the
effective potential (Figure \textbf{1}) corresponding to radial
parameter increases as the mass of BH decreases and angular momentum
increases, consequently, the absorption probability reduces.
\item The height of effective potential (Figures \textbf{2}-\textbf{3})
decreases with a decrease in the coupling parameter and increases
with an increase in charge and frequency parameters implying that
the absorption probability decreases.
\item It is found that the GF (Figure \textbf{4})
corresponding to frequency parameter decreases as the mass of BH
enhances which minimizes the emission rate. The GF increases for the
higher value of angular momentum and vanishes at $l=0$.
\item We have found that the GF (Figures \textbf{5}-\textbf{6}) has an inverse relation
with coupling, charge and radial parameters.
\end{itemize}

The GF for static spherically symmetric BH with linear
electrodynamics was studied in \cite{36}. We have extended this work
for NLED and discussed the effects of physical parameters on GF. The
evaporation rate of a BH can be measured by using the GF as it is
related to the emission of waves through the potential barrier. We
have found that GF with NLED increases more rapidly as compared to
linear electrodynamics. Thus, the presence of a non-linear charge
increases the process of evaporation and can be expected for BH to
die in a short span which is consistent with the literature
\cite{37}. It is worthwhile to mention here that our results reduce
to Reissner-Nordstrom BH for $\alpha=0$ and Schwarzschild BH for
$\alpha=Q=0$. It would be interesting to find GF for rotating
spherically symmetric BH with NLED to reveal the influence of
rotation on the GF.


\begin{thebibliography}{55}

\bibitem{1} Bardeen, J.M.: Proceedings of GR5 (Tiflis, USSR, 1968)174.

\bibitem{2} Kiselev, V.V.: Class. Quantum Grav. \textbf{20}(2003)1187.

\bibitem{3} Hayward, S.: Phys. Rev. Lett. \textbf{96}(2006)031103.

\bibitem{4} Chen, S. and Jing, J.: Class. Quantum Grav. \textbf{22}(2005)4651;
\textbf{23}(2006)6141 ; Gen. Relativ. Gravit. \textbf{39}(2007)1003.

\bibitem{5} Bambi, C. and Modesto, L.: Phys. Lett. B \textbf{721}(2013)329.

\bibitem{6} Xu, Z., Hou, X. and Wang, J.: Class. Quantum Grav. \textbf{35}(2018)115003.

\bibitem{7} Xu, Z. et al.: Eur. Phys. J. C \textbf{78}(2018)513.

\bibitem{8} Hawking, S.W.: Nature \textbf{248}(1974)32; Commun. Math. Phys. \textbf{43}(1975)199.

\bibitem{9} Gubserv, S.S. and Klebanov, I.R.: Phys. Rev. Lett.
\textbf{77}(1996)4491.

\bibitem{10} Maldacena, J.M. and Strominger, A.: Phys. Rev. D
\textbf{55}(1997)861.

\bibitem{11} Klebanov, I.R. and Mathur, S.D.: Nucl. Phys. B
\textbf{500}(1997)115.

\bibitem{12} Kim, W.T. and John, J.: Phys. Lett. B
\textbf{461}(1999)189.

\bibitem{13} Parikh, M.K. and Wilczek, F.: Phys. Rev. Lett.
\textbf{85}(2000)5042.

\bibitem{14} Kerner, R. and Mann, R.B.: Class. Quantum Grav.
\textbf{25}(2008)095014.

\bibitem{15} Ejaz, A. et al.: Phys. Lett. B \textbf{726}(2013)827.

\bibitem{16} Ida, D., Oda, K. and Park, S.C.: Phys. Rev. D \textbf{67}(2003)064025.

\bibitem{17} Creek, S. et al.: Phys. Lett. B \textbf{656}(2007)102.

\bibitem{18} Chen, S., Wang, B. and Su, R.: Phys. Rev. D \textbf{77}(2008)124011.

\bibitem{19} Crispino, L.C.B. et al.: Phys. Rev. D
\textbf{87}(2013)104034.

\bibitem{20} Kanti, P., Pappas, T. and Pappas, N.: Phys. Rev. D \textbf{90}(2014)124077.

\bibitem{21} Jorge, R., de Oliveira, E.S. and Rocha, J.V.: Class. Quantum Grav. \textbf{32}(2015)065008.

\bibitem{22} Toshmatov, B. et al.: Phys. Rev. D \textbf{91}(2015)083008.

\bibitem{23} Ahmad, J. and Saifullah, K.: Eur. Phys. J. C \textbf{77}(2017)885.

\bibitem{24} Dey, S. and Chakrabarti, S.: Eur. Phys. J. C \textbf{79}(2019)504.

\bibitem{25} Sharif, M. and Ama-Tul-Mughani, Q.: Eur. Phys. J. Plus \textbf{134}(2019)616;
Phys. Dark Universe \textbf{27}(2020)100436.

\bibitem{26} Sharif, M. and Shaukat, S.: Ann. Phys. \textbf{436}(2021)168673.

\bibitem{27} Born, M. and Infeld, L.: Proc. Roy. Soc. Lond. A
\textbf{143}(1934)410; \textbf{144}(1934)425.

\bibitem{28} Beato, A.E. and Garcia, A.: Phys. Lett. B \textbf{464}(1999)25.

\bibitem{29} Cai, R.G., Pang D.W. and Wang, A.: Phys. Rev. D \textbf{70}(2004)124034.

\bibitem{30} Bolokhov, S.V., Bronnikov, K.A. and Skvortsova, M.V.: Class.
Quantum Grav. \textbf{29}(2012)245006.

\bibitem{31} Beonnikove, K.A., Dymnikova, I.G. and Galaktinov, E.:
Class. Quantum Grav. \textbf{29}(2012)095025; Yu, S. and Gao, C.:
Int. J. Mod. Phys. A \textbf{29}(2019)2050032.

\bibitem{32} Chowdhury, A. and Banerjee, N.: Phys. Lett. B \textbf{805}(2020)135417.

\bibitem{33} Yu, S. and Gao, C.: Int. J. Mod. Phys. D \textbf{29}(2020)2050032.

\bibitem{34} Flammer, C.: \emph{Spheroidal Wave Functions} (Stanford University Press, 1957).

\bibitem{35} Berti, E., Cardoso, V. and Casals, M.: Phys. Rev. D \textbf{73}(2006)024013; ibid. 109902.

\bibitem{36} Boonserm, P., Ngampitipan, T. and Wongjun, P.: Eur. Phys. J. C \textbf{78}(2018)492.

\bibitem{37} Okyay, M. and {\"O}vg{\"u}n, A.: J. Cosmol. Astropart. Phys. \textbf{2022}(2022)009.
\end{thebibliography}
\end{document}